\documentclass[preprint,12pt,showpacs,aps,amsmath,amssymb,
nofootinbib,superscriptaddress,showkeys]{revtex4}

\usepackage{epsfig}
\usepackage{amsmath}
\usepackage{amssymb}
\usepackage{bbm}
\usepackage{wasysym}
\usepackage{mathrsfs}
\usepackage{graphicx}
\newcommand{\beq}{\begin{equation}}
\newcommand{\eeq}{\end{equation}}
\newcommand{\beqa}{\begin{eqnarray}}
\newcommand{\eeqa}{\end{eqnarray}}

\newcommand{\nn}{\nonumber}
\newcommand{\dd}{\Delta_{31}}
\newcommand{\de}{\Delta_{e}}
\newcommand{\nubar}{\bar{\nu}}
\newcommand{\dmsq}{\Delta m^2}
\newcommand{\dma}{\Delta m^2_{\rm atm}}
\newcommand{\dms}{\Delta m^2_{\odot}}
\newcommand{\thb}{{\theta_b}}
\newcommand{\bb}{\mathbbm}

\begin{document}


\title{CPT violation in long baseline neutrino experiments:\\
a three flavor analysis}

\author{Amol Dighe}
\email{amol@theory.tifr.res.in}
\affiliation{Tata Institute of Fundamental Research, 
Homi Bhabha Road, Colaba, Mumbai 400005, India}

\author{Shamayita Ray}
\email{shamayitar@theory.tifr.res.in}
\affiliation{Tata Institute of Fundamental Research,
Homi Bhabha Road, Colaba, Mumbai 400005, India}


\begin{abstract}
We explore possible signals of CPT violation in neutrinos
in the complete three-flavor framework.
Employing a systematic expansion in small parameters,
we analytically estimate the CPT violating contributions
to the survival probabilities of $\nu_\mu, \bar{\nu}_\mu,
\nu_e$ and $\bar{\nu}_e$.
The results indicate that, in spite of the large number of
CPT violating parameters, only a small number of combinations
are relevant for oscillation experiments.
We identify the combinations that can be constrained 
at the long baseline experiments, and show that their
contribution to the neutrino Hamiltonian can be 
bounded to $\lesssim 10^{-23}$ GeV, by considering the 
NOvA experiment for the muon sector, and neutrino 
factories for the electron sector.
This formalism also allows us to translate the bounds on the 
parameters describing non-standard interactions of neutrinos 
into the bounds on CPT violating quantities.
\end{abstract}



\preprint{TIFR/TH/08-03}

\maketitle

\section{Introduction}\label{intro}

There have been theoretical suggestions that Lorentz invariance 
may not be an exact symmetry \cite{coleman-glashow}. 
In such a case, even if the invariance is broken at a very high 
energy scale (say at the Planck scale in quantum gravity theories), 
the breaking is expected to leave its signature, however small,
at laboratory energies.
Such Lorentz violation may manifest itself as CPT violation.
Indeed, in local field theories, CPT violation
implies Lorentz violation \cite{greenberg-1,greenberg-2}.

In theories with spontaneous CPT violation \cite{kostelecky}, 
the Lagrangian for a fermion to the lowest order in the high
scale can be written as
\beq
{\cal L} = i \bar{\psi} \partial_\mu \gamma^\mu \psi 
-m \bar{\psi} \psi            
- A_\mu \bar{\psi} \gamma^\mu \psi 
- B_\mu \bar{\psi} \gamma_5 \gamma^\mu \psi  \; ,
\label{L-psi}
\eeq
where $A_\mu$ and $B_\mu$ are real numbers.
The terms containing $A_\mu$ and $B_\mu$ are clearly 
Lorentz violating, and give rise to an effective 
contribution to the neutrino Lagrangian that
can be parametrized as 
\beq
{\cal L}_\nu^{CPTV} =  
\bar{\nu}_L^\alpha \, b^{\alpha \beta}_\mu \, 
\gamma^\mu \, \nu_L^{\beta} \; .
\label{L-nu}
\eeq
Here $b_\mu$ are four Hermitian $3\times 3$
matrices corresponding to the four Dirac indices $\mu$,
wherein $\alpha, \beta$ are flavor indices.
Then the effective Hamiltonian for 
ultra-relativistic neutrinos with definite momentum $p$ is
\beqa 
{\bb H} \equiv \frac{{\bb M} {\bb M}^\dagger}{2 p} + {\bb b} \; , 
\label{eff-H}
\eeqa
where ${\bb b} \equiv b_0$ and ${\bb M}$ is the neutrino mass matrix in the 
CPT conserving limit. Following \cite{coleman-glashow},
we choose to work in the preferred frame in which the CMBR 
is isotropic, where the rotational invariance implies no directional
dependence for ${\bb b}$.

The same effective Hamiltonian can also be obtained by considering 
a modified dispersion relation for neutrinos, 
$E^2 = F(p,m)$, in the presence of Lorentz violation.
This dispersion relation may be written, using rotational
invariance in the CMBR frame 
and demanding Lorentz invariance at low energy, as
\cite{mattingly}
\beqa
E^2 = m^2 + p^2 + E_{Pl} f^{(1)}|p| + f^{(2)} p^2 + 
\frac{f^{(3)}}{E_{Pl}} |p|^3 + \cdots \; ,
\label{dispersion}
\eeqa 
where $f^{(n)}$'s are dimensionless quantities.
The Planck energy $E_{Pl}$ is introduced since it is the energy
where Lorentz invariance is expected to be broken in
quantum gravity.
For ultra-relativistic neutrinos with fixed momentum $p$, 
the dispersion relation becomes
\beqa
E = p + \frac{m^2}{2 p} + b + \cdots \; \; ,
\eeqa
such that $b=E_{Pl} f^{(1)}/2$ is the leading CPT violating 
contribution.
Generalizing this to three flavors leads to the same
effective Hamiltonian as in (\ref{eff-H}).

The possible origin of CPT violation in the neutrino sector 
has been studied in the context of  
extra dimensions  \cite{arkani-hamed,huber},
non-factorizable geometry \cite{grossman}, and
non-local causal Lorentz invariant theories \cite{barenboim}.
Bounds on the CPT violating parameters have been 
obtained in many different contexts.
For example, the analyses of neutral meson mixings give
$| m_{K^0} - m_{\overline{K}^0} | \lesssim 10^{-18} 
\; m_{\rm avg}$ \cite{k-kbar}, and
$| m_{B_d^0} - m_{\overline{B_d}^0} | \lesssim 1.6 \cdot 10^{-14} 
\; m_{\rm avg}$ \cite{Bd-Bdbar}, whereas
experiments on anomalous magnetic moment of muon put  
the bound on the anomalous frequency as 
$|\omega_a^{\mu^+} - \omega_a^{\mu^-} | 
\lesssim 10^{-23} \; m_\mu$ \cite{anomalous-g}. 
However, it is difficult to compare these bounds 
directly with the bounds obtained from the neutrino 
sector since we do not have an all-encompassing theory
of CPT violation.

The formalism to analyze CPT violating effects on neutrino
oscillations has been proposed
for the two-flavor case in \cite{coleman-glashow}.
The CPT violating contribution to the Hamiltonian would
change the effective neutrino masses, which in turn would
affect the neutrino oscillation wavelengths. 
Henceforth, for neutrinos we shall use $p \approx E$, 
so that ${\bb M}^2/(2p) \rightarrow {\bb M}^2/(2E)$.
The typical frequency of neutrino oscillations is $\dmsq/(2E)$, 
which can be as small as $10^{-22}$ GeV in the atmospheric 
and long baseline experiments. 
Since the experiments measure the oscillation frequencies to an 
accuracy of $\sim 10 \%$, it may be naively estimated that
neutrino experiments would constrain the CPT violating
parameters to the order $\sim 10^{-23}$ GeV.

After the LSND result \cite{LSND} indicated
three distinct neutrino mass squared differences
when combined with the solar and atmospheric neutrino 
observations,
it was proposed that the CPT violating effects may be 
large enough to make the neutrino and antineutrino
spectra significantly different 
\cite{LSND-Smirnov,Yanagida}.
However, this scenario was found not to be viable when
combined with other neutrino experiments \cite{gouvea-cpt},
and the subsequent observation of oscillations corresponding to
$\dmsq_\odot$ in antineutrinos at KamLAND \cite{kamland}
ruled it out.
If the LSND results are ignored in the light of the
negative results of MiniBooNE \cite{MiniBooNE} that
explore the same parameter space,
CPT violation is not required to explain any neutrino oscillation 
data. 
However, the current uncertainties in the measurements of  
$\dms$ and $\dma$, which are $\sim 9\%$ and $\sim 14\%$ respectively 
\cite{lisi}, allow the possibility of CPT violating effects
in neutrino oscillations, which may be observed or constrained
at the future high precision neutrino oscillation experiments.

CPT violation in neutrino oscillations would manifest itself
in the observation $P(\nu_\alpha \to \nu_\beta) \neq
P(\bar\nu_\beta \to \bar\nu_\alpha)$.
However, when neutrinos propagate through matter, the matter 
effects give rise to ``fake'' CP and CPT violation even 
if the vacuum Hamiltonian is CPT conserving.
These fake effects need to be accounted for while 
searching for CPT violation.
The $\nu_\mu \to \nu_\mu$ channel was explored in this
context in \cite{barger-pakvasa},
where it was pointed out that CPT violating signals could become 
larger due to resonant effects. 
Using a two-flavor analysis, it was shown that 
a long baseline ($L=735$ km) experiment with a typical neutrino factory 
setup can detect a difference of the eigenvalues 
of ${\bb b}$ upto $\delta b \sim 10^{-23}$ GeV.
In \cite{uma-sankar} it was shown that, using the atmospheric
neutrino data at a 50 kt magnetized iron calorimeter 
with 1.2 T magnetic field, $\delta b \sim 3 \times 10^{-23}$ GeV 
should be clearly discernible in 8 years. 
The solar and KamLAND data gives the bound 
$\delta b \lesssim 1.6 \times 10^{-21}$ GeV in \cite{solar-KamLAND}.
Ref.~\cite{bilenky} showed that for a hierarchical neutrino 
mass spectrum, the upper bound for the neutrino-antineutrino mass 
difference that can be achieved in a neutrino factory 
is $|m_3 - \bar{m}_3| \lesssim 1.9 \times 10^{-4}$ eV. 
Global two-flavor analysis of the full atmospheric data and long baseline 
K2K data puts the bound $\delta b \lesssim 5.0 \times 10^{-23}$ GeV 
\cite{garcia-maltoni}. 

All the above analyses have been carried out in the two-flavor
approximation. 
Moreover, it has been assumed in 
\cite{barger-pakvasa} and \cite{uma-sankar} that 
the mixing angles as well as phases of the unitary matrices
that diagonalize ${\bb M}$ and ${\bb b}$ in (\ref{eff-H})
are identical. 
The analysis of \cite{solar-KamLAND} takes the mixing angles
to be identical and considers two specific values of the 
relative phase between the two unitary matrices.
These assumptions have been made solely to
simplify the analytic treatment, and do not have any
physical motivation behind them. 
Ref.~\cite{garcia-maltoni} analyzes the two-flavor case 
in its full generality, putting no extra condition on 
the mixing angles and the relative phase. However,
a three flavor treatment is needed in order to
obtain reliable results, since a two-flavor analysis
cannot account for the CP violating effects that
may interfere with the identification of CPT violation.
The addition of the third (electron) flavor also 
compels one to take care of the matter effects when neutrinos
pass through the Earth.

In this article we consider the possible CPT violating effects 
that appear through (\ref{eff-H}), 
when three-neutrino oscillations are considered in 
their full generality.
We treat the effect of the CPT violating term as a perturbation parametrized 
by a dimensionless auxiliary parameter $\epsilon \equiv 0.1$ and 
express the differences of the eigenvalues of the CPT violating ${\bb b}$ matrix, 
the reactor angle $\theta_{13}$, and the ratio 
$\dmsq_\odot/\dmsq_{atm}$ as some 
power of $\epsilon$ multiplied by ${\cal O}(1)$ numbers, so 
that a systematic expansion in $\epsilon$ can be carried out. 
The survival probabilities of $\nu_\mu$ and $\nu_e$ (and
their antiparticles) can then be written down as 
a power series in $\epsilon$ in a transparent form.
This allows us to identify the combinations of CPT
violating parameters that contribute to these probabilities
to leading order in $\epsilon$.
We compare the signals in the channels
$\nu_\mu \to \nu_\mu$, $\nubar_\mu \to \nubar_\mu$ and 
$\nu_e \to \nu_e$, $\nubar_e \to \nubar_e$
to estimate the extent to which these CPT violating combinations
can be constrained or identified
in future long baseline experiments.

Bounds have been obtained on parameters describing the non-standard
interactions (NSI) of neutrinos with matter 
\cite{maltoni-valle-atm,friedland-lunardini-atm,sacha-davidson,
yasuda-minos,friedland-lunardini-minos,ohlsson-minos},
using both oscillation and non-oscillation experiments.
We explicitly show how to translate these bounds into the
bounds on the CPT violating parameters in our formalism. 
This will allow us to compare and combine the bounds from these
two approaches to restrict new physics in the neutrino sector.

In the paper, Sec.~\ref{sec-General} gives the 
parametrization of the CPT violating part of the effective 
Hamiltonian in flavor basis using the perturbative expansion 
scheme. Sec.~\ref{sec-Pmumu} and Sec.~\ref{sec-Pee} give 
the probability expressions and possible signatures in the 
long baseline experiments in $\mu$ and $e$ channels respectively. 
In Sec.\ref{nsi} we summarize the current constraints on NSI
parameters and translate them to the bounds on CPT violating
quantities.
Sec.~\ref{conclusion} summerizes our results.


\section{Perturbative expansion of the Hamiltonian with
the CPTV term}
\label{sec-General}

In the three neutrino oscillation scheme, $(\nu_e, \nu_\mu, \nu_\tau)$ 
form the flavor basis and $(\nu_1, \nu_2, \nu_3)$ form the mass basis, 
{\it i.e.} the basis in which ${\bb M} {\bb M}^\dagger/(2E)$ is diagonal, 
and these are related by
\beqa
\nu_\alpha = {[\mathbbm{U}_0]}_{\alpha i} \nu_i \; ,
\eeqa
where $\alpha \in \{e, \nu, \tau \}$ and $i \in \{ 1, 2, 3 \}$. 
Let $(\nu^b_{1}, \nu^b_{2}, \nu^b_{3})$ 
be the basis in which $\mathbbm{b}$ is diagonal and let this basis be 
related to the flavor basis as
\beqa
\nu_\alpha = {[\mathbbm{U}_b]}_{\alpha x} \nu^b_{x} \; ,
\eeqa
where $x \in \{ 1, 2, 3 \}$. Both $\mathbbm{U}_m$ and $\mathbbm{U}_b$ 
are unitary matrices.\footnote{
In refs.~\cite{barger-pakvasa,uma-sankar,solar-KamLAND,garcia-maltoni},
${\bb U}_0$ and ${\bb U}_b$ are $2\times 2$ matrices.
In addition, \cite{barger-pakvasa} and 
\cite{uma-sankar} take these two matrices to be identical.}

When neutrinos pass through the matter, the electron neutrinos acquire an 
effective potential $V_e = \sqrt{2} G_F N_e$ due to their charged current 
forward scattering interactions, compared to the other two flavors.
Here $G_F$ is the Fermi constant and  $N_e$ is the number density of 
electrons.
For anti-neutrinos, the sign of $V_e$ is reversed. 
The effective Hamiltonian in the flavor basis is 
\beq
{\bb H}_f \approx \mathbbm{U}_0 \cdot 
\frac{ {\rm diag}(0, \dmsq_{21}, \dmsq_{31})}{2E}  \cdot 
    \mathbbm{U}_0^\dagger
 + {\bb H}_b + {\rm diag}(V_e, 0, 0) \; ,
\label{Hf-1}
\eeq
where
\beq
{\bb H}_b \equiv 
    \mathbbm{U}_b \cdot {\rm diag}(0, b_{21}, b_{31}) \cdot 
    \mathbbm{U}_b^\dagger \; . 
\label{hb-def}
\eeq
Here $b_1$, $b_2$ and $b_3$ are the eigenvalues of $\mathbbm{b}$ and 
$b_{i1} \equiv b_i - b_1$ for $i = 2,3$.
The net spectrum of neutrino mass eigenstates is given by the eigenvalues
of ${\bb H}_f$.
We term the unitary matrix diagonalizing ${\bb H}_f$ as ${\bb U}_f$.


A general $N \times N$ unitary matrix $\mathbbm{U}_N$ is parametrized
by $N (N-1)/2$ independent real quantities (angles) and $N (N+1)/2$ 
independent imaginary quantities (phases). 
In case of the neutrino mixing matrix, $(2N-1)$ phases can be 
absorbed by redefining lepton and neutrino wave-functions.\footnote{
For Majorana neutrinos, only $N$ phases can be absorbed if ${\bb M}$
needs to be kept invariant. However, $N-1$ more phases are
irrelevant when ${\bb M}$ only appears through the combination
${\bb M}{\bb M}^\dagger$.} 
Thus we can parametrize the $3 \times 3$ unitary matrix $\mathbbm{U}_0$ 
by three angles $\theta_{12}$, $\theta_{23}$ and $\theta_{13}$ 
and one phase $\delta_{cp}$.
We write ${\bb U}_0$ in the standard CKM parametrization as 
\beqa
\mathbbm{U}_0 &=& \mathbbm{U}_{23}(\theta_{23},0) \cdot 
\mathbbm{U}_{13}(\theta_{13},\delta_{cp})  \cdot
\mathbbm{U}_{12}(\theta_{12},0) 
\equiv {\bb U}_{CKM} (\{ \theta_{ij} \}; \delta_{cp} ) \; ,
\label{u-ckm}
\eeqa
where  ${\bb U}_{ij}(\theta_{ij},\delta_{ij})$ is the complex rotation
matrix in the $i$--$j$ plane, whose elements
$[{\bb U}_{ij} ] _{pq}$ are defined as
\beq
[{\bb U}_{ij}(\theta,\delta)] _{pq} = \left\{
\begin{array}{ll}
\cos \theta & p=q=i \mbox{ or } p=q=j \\
1 & p=q \neq i \mbox{ and } p=q \neq j \\
\sin \theta e^{-i\delta} & p=i \mbox{ and } q=j \\
- \sin \theta e^{i\delta} & p=j \mbox{ and } q=i \\
0 & {\rm otherwise}  \; .\\
\end{array}
\right.
\label{u-elements}
\eeq

Once we have redefined the phases of the lepton and neutrino 
wavefunctions to get ${\bb U}_0$ in the form (\ref{u-ckm}),
the basis of neutrino flavor eigenstates is completely
defined. 
The matrix $\mathbbm{U}_b$ then needs three angles ${\thb}_{12}$, 
${\thb}_{23}$, $\thb_{13}$ and six phases for a complete
parametrization, given by
\beqa
\mathbbm{U}_b(\{\thb_{ij}\}; \{\phi_{bi}\};\{\alpha_{bi}\};\delta_{b})
& = & {\rm diag}(1, e^{i\phi_{b2}}, e^{i \phi_{b3}}) \cdot
\mathbbm{U}_{CKM}(\{ \thb_{ij}\}; \delta_b) \cdot \nonumber \\
& & \phantom{space}
{\rm diag}(e^{i \alpha_{b1}}, e^{i \alpha_{b2}}, e^{i \alpha_{b3}}) \; ,
\eeqa
where $\alpha_{b1}$, $\alpha_{b2}$, $\alpha_{b3}$ 
are the Majorana phases and will not 
have any contribution to ${\bb H}_f$ through ${\bb H}_b$.
Hence ${\bb U}_f$ may be written in term of a total of 
six mixing angles $(\theta_{12}, \theta_{23}, \theta_{13},
\thb_{12}, \thb_{23}, \thb_{13})$ and four phases
$(\delta_{cp}, \delta_b, \phi_{b2}, \phi_{b3})$.

Present limits on CPT violation in the neutrino sector
\cite{barger-pakvasa,uma-sankar,solar-KamLAND,garcia-maltoni} 
arise from the limits on the neutrino
oscillation wavelength, which in the two-flavor case gets
modified as $\dmsq_{atm}/(2E) \to \dmsq_{atm}/(2E) + \delta b$.
The bound on $\delta b$ is therefore governed by the 
uncertainty in $\dmsq_{atm}$, which is $\sim 10\%$.
Motivated by this, we assume that 
\beq
b_{21}, b_{31} \lesssim 0.1 \times \dmsq_{atm}/(2 E_0) \approx
0.13 \times 10^{-21} \, {\rm GeV}^2/ E_0 \; , 
\label{S_E0-def}
\eeq
where $E_0$ is the
typical energy scale of the experiment.
This may be parametrized by introducing two auxiliary quantities
$\epsilon \equiv 0.1$ and 
$S_{E_0} \equiv 10^{-21} \, {\rm GeV}^2/E_0$
such that
\beq
b_{21} \equiv \epsilon \beta_{21} S_{E_0} \; , \quad
b_{31} \equiv \epsilon \beta_{31} S_{E_0} \; .
\label{approx-b}
\eeq
Clearly, $\beta_{21}, \beta_{31}$ are numbers of ${\cal O}(1)$ 
or smaller.
The mixing angle $\theta_{13}$ and the ratio $\dmsq_{21}/\dmsq_{31}$
are small quantities, and may be expressed in terms of powers of
$\epsilon$ as
\beq
\theta_{13} \equiv \epsilon \chi_{13} \; , \phantom{space}
\dmsq_{21}/\dmsq_{31} \equiv \epsilon^2 \zeta \; .
\label{approx-standard}
\eeq 
The current bounds on the mixing angles and mass squared differences 
\cite{lisi} set 
\beq
\chi_{13} < 1.8 \; ,  \phantom{space} \zeta \sim 3.0 \; .
\eeq
The sign of $\zeta$ is positive (negative) for normal (inverted) mass 
ordering of neutrinos.  


Using the formal representation of $\theta_{13}$, $\dmsq_{21}$ and the 
CPT violating parameters in terms of powers of $\epsilon$ as given in 
eq.~(\ref{approx-b}) and eq.~(\ref{approx-standard}), 
the Hamiltonian ${\bb H}_f$ can be expanded formally 
in powers of $\epsilon$ as
\beq
{\bb H}_f = \frac{\dmsq_{31}}{2E} \left[ {\bb H}^{(0)}_f + 
\epsilon {\bb H}^{(1)}_f + 
\epsilon^2 {\bb H}^{(2)}_f 
+ {\cal O}(\epsilon^3) \right] \; ,
\eeq
where ${\bb H}^{(0,1,2)}_f$ are functions of all the mixing angles, phases, 
mass squared differences, and eigenvalues of $\mathbbm{b}$. 
All the elements of ${\bb H}^{(0,1,2)}_f$ are of ${\cal O}(1)$ 
or smaller, and ${\bb H}^{(0)}_f$ has nondegenerate eigenvalues.
The techniques of time independent nondegenerate perturbation theory 
can therefore be used to calculate the eigenvalues and eigenvectors
of ${\bb H}_f$ upto the required order in $\epsilon$.
These can be further used to calculate the neutrino flavor
survival or conversion probabilities when neutrinos travel
through matter with a constant density:
\beq
P_{\alpha\beta} \equiv P(\nu_\alpha \to \nu_\beta) =
\left|
\sum_{i} [{\bb U}_f]_{\alpha i} [{\bb U}_f]_{\beta i}^\ast
\exp \left(-i \frac{\widetilde{m}_i^2 L}{2E} \right)
\right|^2 \; ,
\label{p-ab}
\eeq
where $\widetilde{m}_i^2/(2E)$ are the eigenvalues of ${\bb H}_f$.
This approximation is valid for neutrino propagation
inside the Earth as long as neutrino trajectories do not 
pass through the core, and neutrino energy is not close to the 
$\theta_{13}$ resonance energy in the Earth matter.


\section{CPT violation in $P_{\mu\mu}$ and signatures at NOvA}
\label{sec-Pmumu}

The survival
probability of muon neutrinos of energy $E$, after traversing 
a distance $L$ through the Earth is given as
\beqa
P_{\mu \mu} &=& 1 - \sin^2{2 \theta_{23}} \sin^2{\dd} \nn \\
&+& \epsilon \Bigl( {C}_1 \Delta_0 \sin^2{2 \theta_{23}} 
 \sin{2 \dd} - {C}_2 \frac{\Delta_0}{\dd} \sin{4\theta_{23}} 
 \sin^2{\dd} \Bigr) + {\cal O}(\epsilon^2) \; ,
\label{pmumu}
\eeqa
where we define the dimensionless quantities
$\dd \equiv \dmsq_{31} L/(4E)$ and
$\Delta_0 \equiv \dmsq_{31} L/(4E_0)$.
The first two terms in (\ref{pmumu}) are CPT conserving, and 
describe oscillations with frequency governed by $\dmsq_{atm}$.
The subleading contribution at ${\cal O}(\epsilon)$ is
CPT violating.
The quantities ${C}_{1,2}$ are given by
\beqa
{C}_1 &=& {B}_1 \cos{2\theta_{23}} -{B}_2 \sin{2\theta_{23}} \; , \nn \\
{C}_2 &=& {B}_1 \sin{2\theta_{23}} +{B}_2 \cos{2\theta_{23}} \; , 
\label{c-def}
\eeqa
where
\beqa
B_1 = ( \mathbbm{H}_{b22} - \mathbbm{H}_{b33} ) / (\epsilon S_{E_0}) \; , 
\quad  B_2 = 2 {\rm Re} (\mathbbm{H}_{b23}) /  (\epsilon S_{E_0}) \; .
\label{B-Hb}
\eeqa
The quantities $B_1$ and $B_2$ depend only on ${\bb b}$ and ${\bb U}_b$ as
\beqa
{B}_1 &=& \left[ X \cos{2 \thb_{23}} - Y \sin{2 \thb_{23}} 
\cos{\delta_b} \right]  \label{B1} \; , \\
{B}_2 &=&  -\left[ X \sin{2 \thb_{23}} 
\cos{\delta \phi} + Y \cos{2 \thb_{23}}\cos{\delta \phi} \cos{\delta_b} + 
Y \sin{\delta \phi} \sin{\delta_b}  \right] 
\label{B2} \; , 
\label{b-def}
\eeqa
wherein
\beqa
X&=&\beta_{21} \cos^2{\thb_{12}}-\beta_{31} \cos^2{\thb_{13}}
-\beta_{21} \sin^2{\thb_{12}}\sin^2{\thb_{13}} \; ,\\
Y &=& \beta_{21}  \sin{2 \thb_{12}} \sin{\thb_{13}}\; .
\label{xy-def}
\eeqa
The phases $\phi_{bi}$ only appear through the combination 
$\delta \phi = {\phi_b}_2 -{\phi_b}_3$.
The corresponding quantity $P_{\bar\mu \bar\mu}$ for antineutrinos
can be obtained simply with the substitution $\epsilon \to 
-\epsilon$ in (\ref{pmumu}).
The terms involving matter effects as well as CP violation are
suppressed due to $\theta_{13}$ and $\dms$, and appear only
at ${\cal O}(\epsilon^2)$ and ${\cal O}(\epsilon^3)$ respectively.
CPT violation can thus be cleanly extracted from the asymmetry
\beq
P_{\mu\mu}-P_{\bar\mu \bar\mu}
= 2 \epsilon \Bigl( {C}_1 \Delta_0 \sin^2{2 \theta_{23}} 
 \sin{2 \dd} - {C}_2 \frac{\Delta_0}{\dd} \sin{4\theta_{23}} 
 \sin^2{\dd} \Bigr) + {\cal O}(\epsilon^2) 
\label{asym-mu}
\eeq
if it is indeed of the 
magnitude allowed by the current bounds.
However, one needs to be away from the $\theta_{13}$ resonance,
which for the Earth matter density occurs for 
$E_{res} \approx 5$--$10$ GeV, since the enhanced value of $\theta_{13}$
makes the expansion in powers of $\epsilon$ invalid.

Eq.~(\ref{pmumu}) demonstrates that, though the CPT violating parameter
space consists of three angles, three phases and two eigenvalue 
differences $\beta_{21}$, $\beta_{31}$, the effective 
CPT violating contribution to the probability $P_{\mu \mu}$ is much simpler and 
depends only on two combinations of these parameters, ${B}_{1}$ and 
${B}_{2}$, to leading order. 
A consequence of this result is that measurements in the muon channel
can only put bounds on the two effective parameters $B_1$ and $B_2$,  
and not separately on the angles, phases or eigenvalue differences. 

Since $\theta_{23} \approx \pi/4$, the $C_1$ term in (\ref{asym-mu})
dominates over the other. Accounting for the $1/L^2$ fall-off of
the neutrino flux, the signal due to this term is optimized 
when $L$ takes its minimum value that is able to satisfy
$\sin 2\Delta_{31} \approx 1$. This calls for a (relatively) low
energy experiment with $L/E \sim 240$ km/GeV.
The NOvA experiment \cite{nova-1,nova-2} with its $L =812$ km baseline
and the NuMI beam energy $E \approx 0.5$--$4.0$ GeV satisfies these
criteria, and hence is well suited to look for
CPT violation. 
The energy range of NOvA is completely below the $\theta_{13}$
resonance energy, so the contamination from CP violating $\theta_{13}$
contributions to $P_{\mu\mu} - P_{\bar\mu \bar\mu}$ is minimal.
We take $E_0$ for NOvA to be 1 GeV, so that 
$S_{E_0} = 10^{-21}$ GeV.

\begin{figure}
\epsfig{file=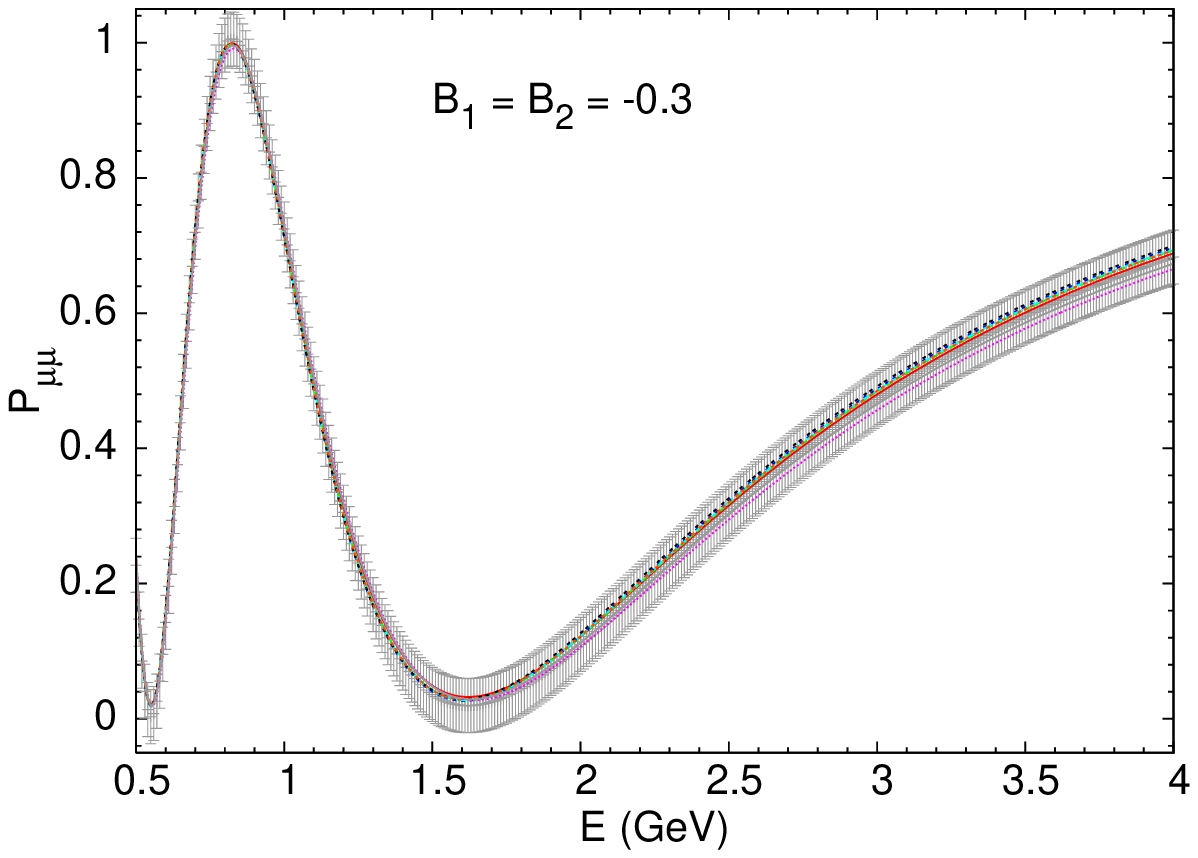,width=7.5cm}
\epsfig{file=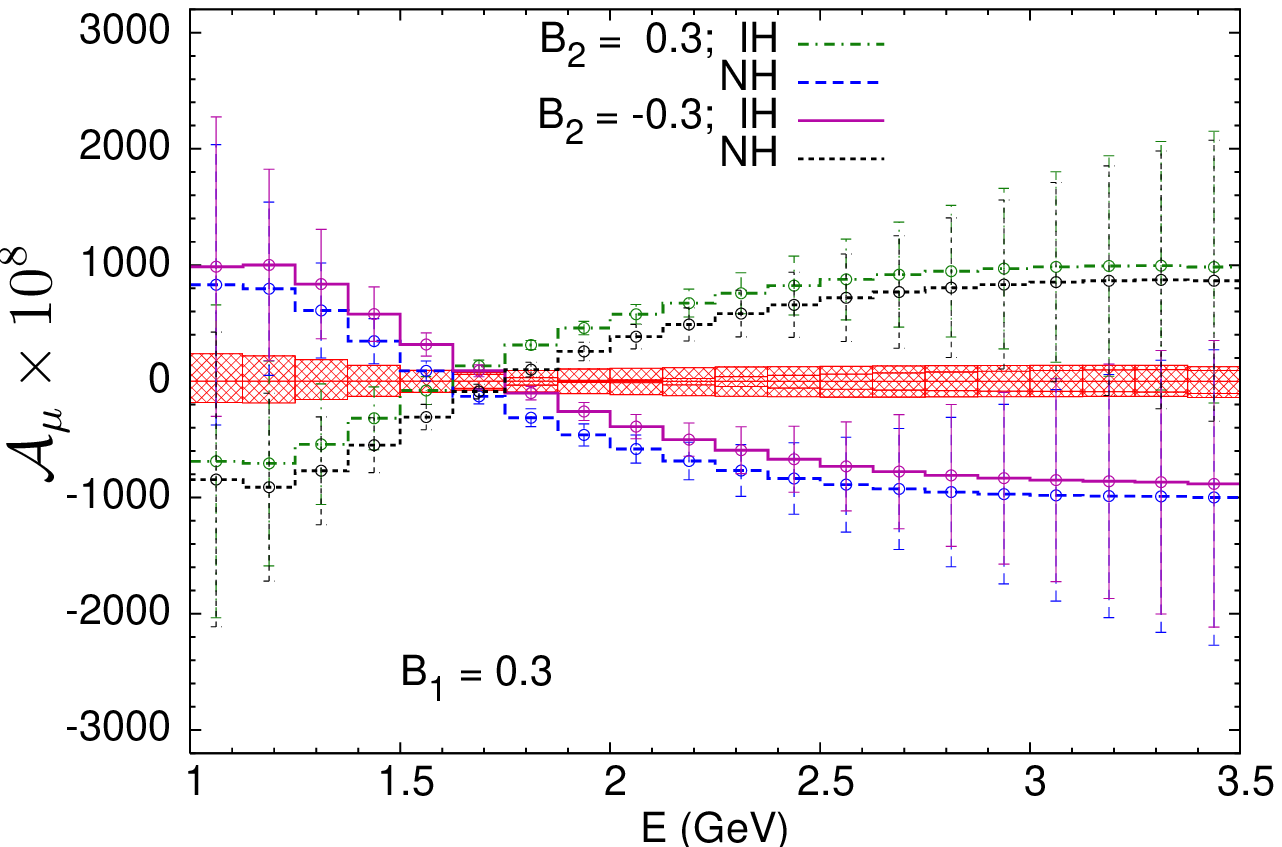,width=7.5cm}
\caption{The left panel compares the analytical and numerical
results for eight randomly chosen combinations of CPT violating
parameters (thin lines) that correspond to 
$B_1 = B_2 = -0.3$. 
The gray (shaded) band is the analytic expression plotted with 
an energy independent error of $\pm 0.04$.
We choose normal hierarchy, $\dms = 7.92 \times 10^{-5}$ eV$^2$, 
$\theta_{12} = 34.08^\circ$, $\dma = 2.6 \times 10^{-3}$ eV$^2$, 
$\theta_{23} = 42.13^\circ$, $\theta_{13} = 0.089$ and $\delta_{cp} = 0$. 
The right panel shows ${\cal A}_\mu(E)$
for 4 years of running at NOvA with each $\mu^+$ and $\mu^-$, 
with an incident flux of $10^{21}$ pot yr$^{-1}$.
The errors shown are only statistical.
The central red (hashed) band shows the contribution in absence of 
CPT violation when the parameters are varied over their current 
$2 \sigma$ ranges.
\label{Pmumu-fig1}}
\end{figure}

We demonstrate the validity and limitations of the analytic expression 
(\ref{pmumu}) in the left panel of Fig.~\ref{Pmumu-fig1}, 
where $P_{\mu \mu}$ is plotted as a 
function of energy for the NOvA baseline for the current best-fit 
values of $\dms$, $\dma$, $\theta_{12}$, $\theta_{23}$ and 
$\theta_{13}$ over the NOvA energy range.
We choose normal mass ordering and $B_1$, $B_2$ with opposite signs, 
which is observed to be one of the worst case situations while comparing
analytical results with numerical ones.
We choose ${B}_1 = {B}_2 = -0.3$ and take
eight randomly chosen sets of CPT violating parameters that correspond to
these values of $B_1$ and $B_2$.
The plot shows that an energy independent error of $\pm 0.04$ 
in $P_{\mu \mu}$ can account for the error due to neglecting terms of 
${\cal O}(\epsilon^2)$ or 
higher, in the whole energy regime of interest.\footnote{
A part of this systematic error should be cancelled out when we
concentrate on $P_{\mu\mu}-P_{\bar\mu \bar\mu}$, however we
choose to use a more conservative estimate of errors.}

We choose a typical NOvA setup \cite{nova-1,nova-2}, with the NuMI beam 
directed towards a $0.5$ kt ``near''  detector placed $1$ km away, 
and a $25$ kt ``far'' detector at a distance of $812$ km. 
The detector is assumed to be able to identify lepton charges. 
The neutrino propagation through the Earth is implemented using a 
5-density model of the Earth, where the density of each layer has been 
taken to be the average of the densities encountered by the neutrinos 
along their path in that layer with the PREM profile \cite{prem}. 
We take care of the detector characteristics 
using the General Long Baseline Experiment Simulator
(GLoBES) \cite{globes-1,globes-2}. The cross-sections used are 
taken from \cite{xsec-1,xsec-2}, and the simulation 
includes an energy resolution of $\sigma_E = 10\% \; \sqrt{E}$,
an overall detection efficiency of 80\% for all charged leptons,
as well as additional energy dependent post-efficiencies 
that are taken care of bin-by-bin {\it a la} GLoBES.
We assume perfect lepton charge identification, and 
neglect any error due to wrong sign leptons produced from the 
oscillations of the antiparticles.

In the right panel of Fig.~\ref{Pmumu-fig1}, we plot the asymmetry 
\beqa
{\cal A}_{\mu}(E) \equiv
\frac{N_\mu^{\rm far}(E)}{N_\mu^{\rm near}(E)} - 
\frac{{\overline{N}}_{\mu}^{\rm far}(E)}{{\overline{N}}_{\mu}^{\rm near}(E)} \; ,   
\eeqa
where $N_\ell$ (${\overline{N}}_{\ell}$) is the number of $\ell^-$ 
($\ell^+$) observed at the near or far detector. Here the events 
observed in the near detector act as a normalizing factor, and 
help in canceling out the systematic errors due to fluxes, 
cross sections and efficiencies in each energy bin. 
Note that modulo these factors, ${\cal A}_\mu$ is equivalent to
$(P_{\mu\mu}-P_{\bar\mu \bar\mu})$ multiplied by a geometric factor of 
$(L_{near}/L_{far})^2$.
For plotting, 
we have considered a running time of $4$ years with each of $\mu^+$ and 
$\mu^-$, with an incident flux of $10^{21}$ pot (protons on target)
per year.

The right panel of Fig.~\ref{Pmumu-fig1} illustrates salient features of the CPT
violating contribution to ${\cal A}_\mu(E)$. 
The central band corresponds to possible 
signals in the absence of any CPT violating contributions, 
where we have varied $\theta_{23}$, $\theta_{13}$ and $\delta_{cp}$ 
over the currently allowed $2\sigma$ ranges and have 
allowed for normal as well as inverted mass ordering.
We fix $\dms$, $\dma$ and $\theta_{12}$ at their current best fit 
values, since variation with these parameters is not expected to be 
significant.
For illustrating the signal in the presence of CPT violation,
we choose $\delta_{cp} = 0$, $|B_{1,2}| = 0.3$, 
and fix $\theta_{23}$ and $\theta_{13}$ at their best fit 
values.
The figure shows that ${\cal A}_\mu$ depends on both the magnitude 
and relative sign of ${B}_{1}$, $B_2$ and also on the 
mass ordering. 
It can be shown that the effect of changing sign of 
${B}_{1,2}$ is the same as changing the mass ordering, 
as expected from eq.~(\ref{asym-mu}) when $\theta_{23} \approx \pi/4$.

\begin{figure}
\epsfig{file=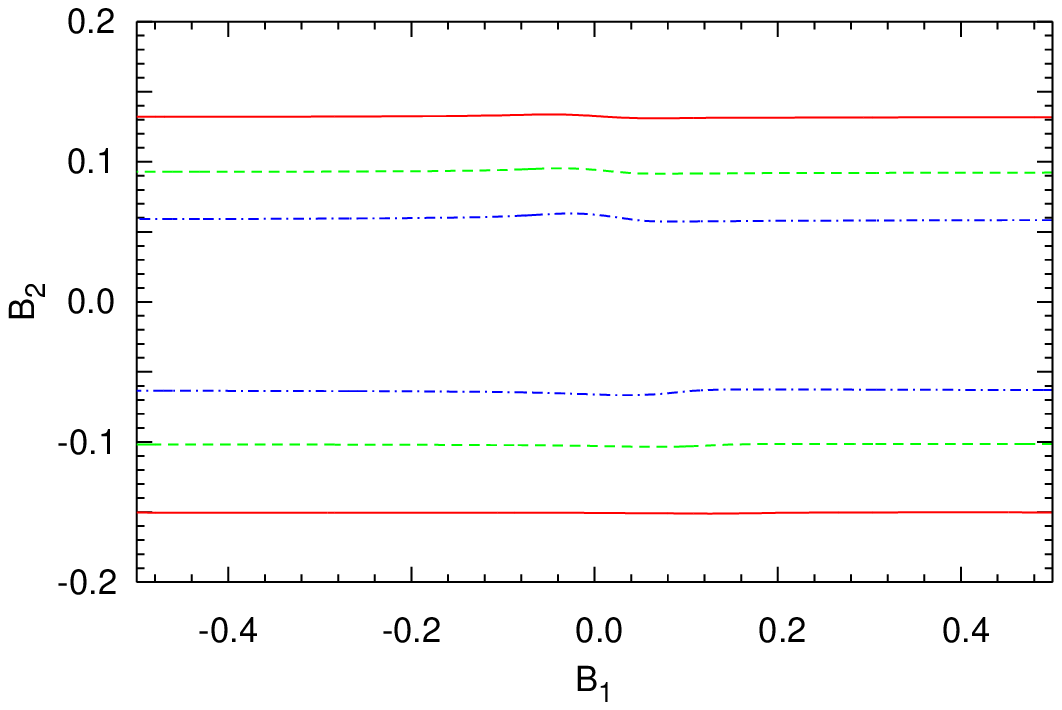,width=7.5cm}
\epsfig{file=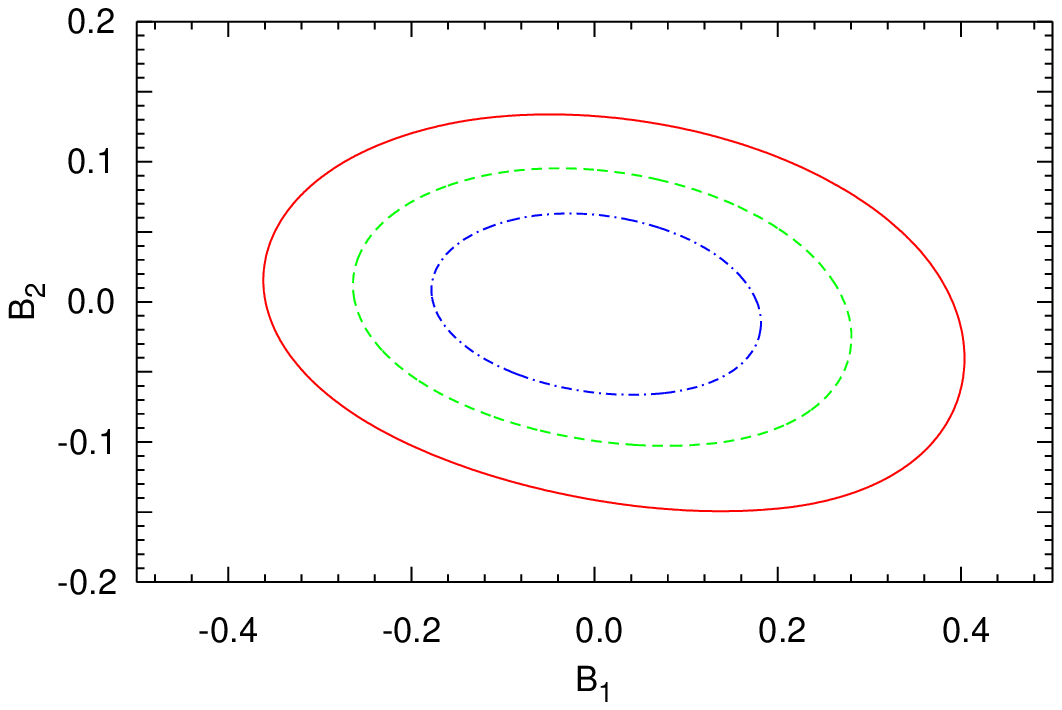,width=7.5cm}
\caption{Confidence level contours in the $B_1$--$B_2$ plane. The red (solid), 
green (dashed) and blue (dash-dotted) curves give the $3 \sigma$, 
$2 \sigma$ and $1 \sigma$ contours respectively. 
An energy independent error of $\pm 0.04$ on $P_{\mu \mu}$ 
has been taken into account.
Both the figures use 
$\theta_{12} = 34.08^\circ$, $\dma = 2.6 \times 10^{-3}$ eV$^2$, 
$\theta_{13} = 0.089$ and $B_1 = B_2 = 0$ as the input values.
The left figure marginalizes over 
$\theta_{23}, \dmsq_{atm}$ and $\delta_{cp}$. 
The right figure uses $\theta_{23} = 42.13^\circ$. 
\label{contour}}
\end{figure}

In order to estimate the possibility of identifying the CPT violating 
contributions from the experimental signals, with the current uncertainties 
in the standard three neutrino oscillation parameters, we display the 
confidence level contours in Fig.~\ref{contour}. 
In the left panel of Fig.~\ref{contour}, we have marginalized over 
all the standard neutrino 
oscillation parameters.
It shows that $B_2$ can be bounded from NOvA observations 
to the extent
\beq
|B_2| \lesssim 0.1 \quad  (2\sigma) \ \; .
\label{b2lim}
\eeq

The data are relatively insensitive to $B_1$. 
This is expected from the analytic expression in eq.~(\ref{c-def}) 
and (\ref{asym-mu}): since $\theta_{23} \approx \pi/4$, 
the terms containing $B_1$ are highly suppressed. 
However, if $\theta_{23}$ is known accurately and differs from $\pi/4$,
the sensitivity to $B_1$ is restored.
This becomes clear from the right panel of Fig.~\ref{contour}, where we 
have marginalized over all parameters except $\theta_{23}$, 
keeping $\theta_{23}$ fixed at a non-maximal value of $42.13^\circ$.
In such a case, $B_1$ can be constrained to
\beqa
|B_1| \lesssim 0.28 \quad (2\sigma) \ \; .
\label{b1lim}
\eeqa
Accurate measurement of the deviation of $\theta_{23}$ from its maximal
value \cite{maximal-t23} is essential for the above bound on $B_1$.
The bound on $B_2$, however, does not depend on the improvement in
the measurement of any other quantity.
A similar analysis performed for inverted hierarchy gives
virtually identical results.

The limits obtained in (\ref{b2lim}) and (\ref{b1lim})
are bounds on specific
combinations of elements of ${\bb H}_b$. They imply
\begin{eqnarray}
\mathbbm{H}_{b22} - \mathbbm{H}_{b33} & = & \epsilon |B_1| S_{E_0}
\lesssim 10^{-23} ~{\rm GeV} \; , \\
2 {\rm Re}(\mathbbm{H}_{b23}) & = & \epsilon |B_2| S_{E_0}
\lesssim 10^{-23} ~{\rm GeV} \; .
\label{bound-B-CPTV}
\end{eqnarray}
In the limit 
${\theta_{b}}_{23}=\pi/4, \beta_{21}={\theta_{b}}_{13}=\delta \phi= 0$,
the quantity $B_2$ in fact reduces to $\delta b$, 
the quantity bounded in the two-flavor analysis 
\cite{barger-pakvasa,uma-sankar}.
Our three
neutrino analysis thus identifies the quantity that can be 
constrained, and also demonstrates that the constraints can be
quantified in a clean manner at a low energy
long baseline experiment like NOvA.


\section{CPT violation in $P_{ee}$ and signatures at a
neutrino factory}
\label{sec-Pee}

The survival probability for an electron neutrino travelling 
through a uniform matter density, in the presence of
CPT violation, is given by 
\beqa
P_{ee} &=& 1 - 4 \epsilon^2 \chi_{13}^2 \left( 
\frac{\dd}{\de - \dd} \right)^2 
\sin^2{(\de - \dd)} \nn \\
&-& 4 \epsilon^2 \chi_{13} \
\left[ {\rm Re}\left( S e^{i \delta_{cp}} \right) \right]
\frac{\dd \Delta_0}{ (\de - \dd)^2}
\sin^2{(\de - \dd)} 
\nn \\
&-&  \epsilon^2 
P_1 \cos{\delta_{cp}} 
\frac{\Delta_0^2}{4 \Delta_e^2} 
\frac{2 \Delta_e^2 - 2 \Delta_e \Delta_{31} + \Delta_{31}^2}{
(\Delta_e - \Delta_{31})^2 } 
\nn \\
&+&   \epsilon^2 \ 
( P_2 \cos{2 \theta_{23}} + P_3 \sin{2 \theta_{23}} )
\cos{\delta_{cp}} 
\frac{\Delta_0^2}{4 \Delta_e^2} 
\frac{\Delta_{31} ( -2 \Delta_e + \Delta_{31})}
{(\Delta_e - \Delta_{31})^2 } 
\nn \\
&+& \epsilon^2 \biggl[ P_4 \frac{\Delta_0^2}{2 \Delta_e^2} 
\cos{2\de} - 
|S|^2 \frac{2 \Delta_0^2}{(\Delta_e - \Delta_{31})^2} 
\cos{(2 \de - 2 \dd)}\biggr] + {\cal O}(\epsilon^3) \; ,
\label{pee}
\eeqa
where 
$\Delta_e \equiv V_e L/2$ and recall that $\theta_{13} = \epsilon \chi_{13}$.
The CPT violating quantities appearing in (\ref{pee}) can be expressed in 
terms of two complex quantities $Q \equiv \mathbbm{H}_{b13}$ and 
$R \equiv \mathbbm{H}_{b12}$:
\beqa
Q \equiv Q_1 + i Q_2 &=& - \frac{1}{2} \cos{\thb_{13}}  e^{-i {\phi_b}_3}
\biggl\{ \beta_{21} \sin{2 \thb_{12}} \sin{\thb_{23}} \biggr. \nn \\
&& \biggl. - 2 \left( \beta_{31} - \beta_{21} \sin^2{\thb_{12}} \right) \sin{\thb_{13}} 
\cos{\thb_{23}} e^{-i \delta_b}  \biggr\} \; , \\
R \equiv  R_1 + i R_2 &=& \frac{1}{2} \cos{\thb_{13}}  e^{-i {\phi_b}_2}
\biggl\{ \beta_{21} \sin{2 \thb_{12}} \cos{\thb_{23}} \nn \\
&& \biggl. + 2 \left( \beta_{31} - \beta_{21} \sin^2{\thb_{12}} \right) \sin{\thb_{13}} 
\sin{\thb_{23}} e^{-i \delta_b}  \biggr\} \; ,
\eeqa
where $Q_i$ and $R_i$ are real numbers.
In terms of $Q$ and $R$, 
the CPT violating parameters $S$ and $P_{1,2,3,4}$ 
in (\ref{pee}) may be written as
\beqa
S &=&  Q \cos{\theta_{23}} + R \sin{\theta_{23}} \; , \\
P_1 &=& |Q|^2 + |R|^2 \; , \quad P_2 = |Q|^2 - |R|^2 \; , \quad 
P_3 = 2{\rm Re} (Q R^\ast)\; , \quad P_4 = P_1 - |S|^2 \; . 
\eeqa

In eq.~(\ref{pee}), the first two terms are the expression for 
$P_{ee}$ with a non-zero $\theta_{13}$ when CPT is conserved, 
while all the other terms are CPT violating contributions. 
There are no ${\cal O}(\epsilon)$ terms.
Both the $\theta_{13}$ correction as well as the CPT violating contributions 
appear at ${\cal O}(\epsilon^2)$.
These also include terms that get contributions from both 
$\theta_{13}$ and CPT violating parameters. 
The probability $P_{ee}$ depends only on two complex combinations $Q$ and $R$ 
of CPT violating parameters, so this channel can 
put bounds only on these two parameters.

Note that the coefficients of ${\cal O}(\epsilon^2)$ in (\ref{pee}) 
contain terms proportional to $(\Delta_{31}/\Delta_e)^2$,
which should be small for the $\epsilon$-expansion to be
under control. 
Therefore, the expression (\ref{pee}) is valid only when 
$\Delta_e \gtrsim \Delta_{31}$, which happens at energies above
the $\theta_{13}$ resonance energy. 
In order to get significant effects at large energies, 
one also needs long baselines.
Both these conditions would be  
satisfied at a neutrino factory with an energy range 
$10$--$50$ GeV and a baseline $\sim 3000$ km. 
If we restrict ourselves to energies well above the $\theta_{13}$
resonance energy $\approx 5$--$10$ GeV, even the CPT conserving
$\theta_{13}^2$ contribution is suppressed, so that the CPT violating
contribution can be more cleanly identified.
For the neutrino factory,  we can set the typical energy $E_0 = 10$ GeV, 
so that $S_{E_0} = 10^{-22}$ GeV. 

\begin{figure}
\epsfig{width=7.5cm,file=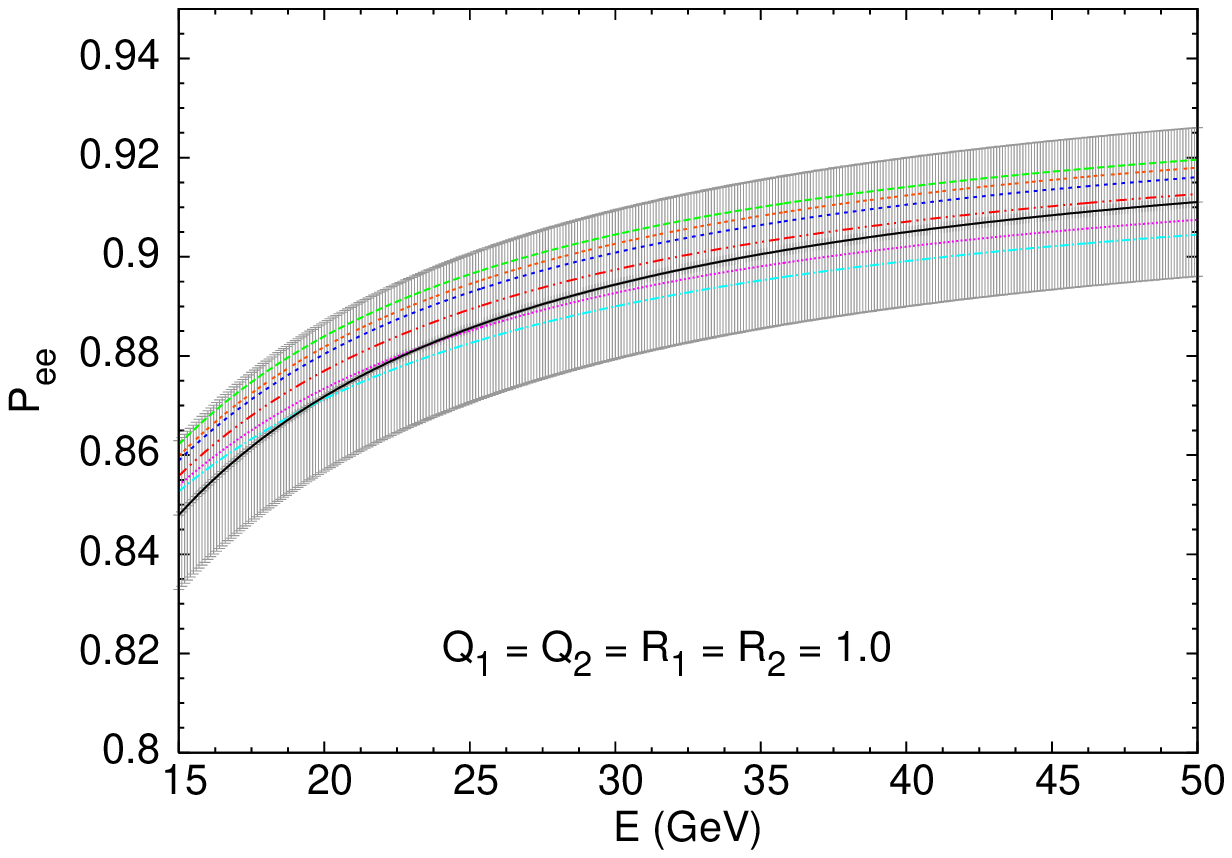}
\epsfig{width=7.5cm,file=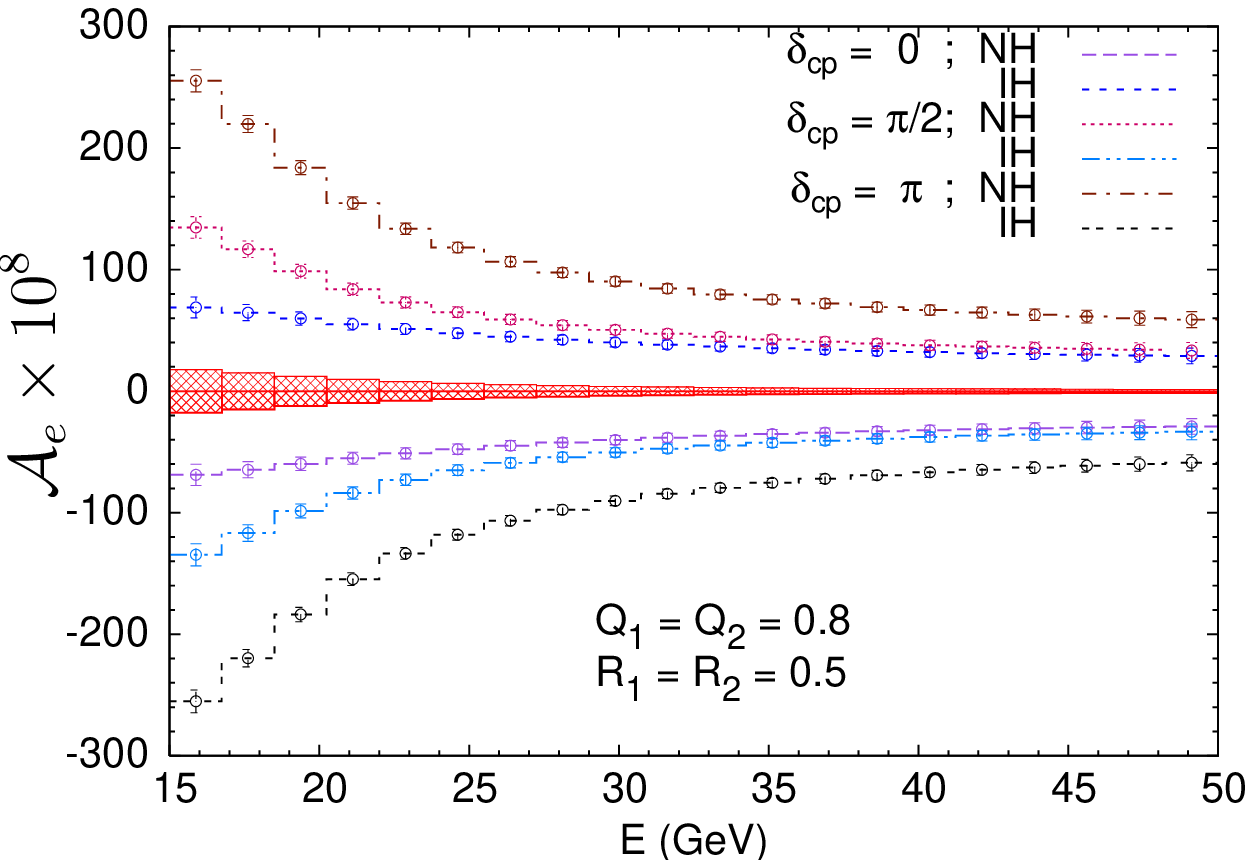}
\caption{
In the left panel, 
the black (solid) line is the analytic expression, and the 
gray (shaded) band corresponds 
to the analytic value with an error $\pm 0.015$. The lines in six 
different colors (symbols) are for six 
random sets of CPT violating parameters with $Q_1$, $Q_2$, $R_1$, $R_2$ fixed 
at $1.0$.
We choose normal ordering, and the values of $\dms$, $\theta_{12}$, $\dma$, 
$\theta_{23}$, $\theta_{13}$ and $\delta_{cp}$ 
the same as that in Fig.~\ref{Pmumu-fig1}. 
The neutrinos traverse through the Earth for $L = 3000$ km before being 
detected.
In the right panel,
the central band shows the contribution in absence of 
CPT violation when the parameters are varied over their current 
$2 \sigma$ ranges. The counts are for 2 years of running of neutrino factory 
with each of $e^+$ and $e^-$. The errors are only statistical.
}
\label{pee-plot}
\end{figure}

In Fig.~\ref{pee-plot} we demonstrate the validity and limitations 
of the analytic probability expression (\ref{pee}), where we choose the 
mixing parameters $\dms$, $\dma$, $\theta_{12}$, $\theta_{23}$, $\theta_{13}$ 
to have their best-fit values \cite{lisi}. We choose normal mass 
hierarchy, and fix $\delta_{cp} = 0$. 
This is observed to be one of the worst case situations 
while comparing the analytical expressions with the numerical ones.
For the CPT violating part we choose $Q_1 = Q_2 = R_1 = R_2 = 1.0$,
and six random choices of the elements of ${\bb b}$ that map to
these values of $Q_i$s and $R_i$s.
It is seen that an energy independent error of $\pm 0.015$ on
$P_{ee}$ can account for the error due to neglecting higher order
terms in $\epsilon$ over the whole energy range of interest.

To demonstrate the capability of a typical neutrino factory setup 
for identifying the CPT violating contributions, we define the asymmetry 
\beqa
{\cal A}_{e}(E) \equiv
\frac{N_e^{\rm far}(E)}{N_e^{\rm near}(E)} - 
\frac{{\overline{N}}_{e}^{\rm far}(E)}{{\overline{N}}_{e}^{\rm near}(E)} \; 
\eeqa
in a typical neutrino factory
setup \cite{nufact-setup} with a 50 GeV muon beam 
directed to a 0.5 kt ``near'' detector 1 km away, and a 
50 kt ``far'' detector 3000 km away.
The detectors are assumed to be
capable of identifying lepton charges. 
The number of useful muons in the storage ring
is taken to be $1.066 \times 10^{21}$, which corresponds to
approximately two years of running with $\mu^-$ and 
$\mu^+$ each at the neutrino factory, using the 
NuFact-II parameters in \cite{nufactII}. The simulation 
includes an energy resolution of $\sigma_E/E = 15\%$,
and an overall detection efficiency of 75\% for all charged leptons.
Earth matter 
effects, interaction cross-sections and post-efficiencies are taken care of 
in the same way as was done in the case of NOvA. 
We assume perfect lepton charge identification, and 
neglect any error due to wrong sign leptons produced 
from the oscillations of the antiparticles. 
GLoBES is used to get the energy variation of the asymmetry 
${\cal A}_e (E)$ as shown in the right panel of
Fig.~\ref{pee-plot}.
The figure indicates that it will be possible 
to discern the CPT violating contributions from the background of CPT 
conserving contributions if, for example, 
$Q_i = 0.8$ and $R_i = 0.5$.
Note that ${\cal A}_e$ is approximately equivalent to
$(P_{ee}-P_{\bar{e} \bar{e}})$ multiplied by a geometric factor of 
$(L_{near}/L_{far})^2$.

\begin{figure}[htp]
\epsfig{width=7.0cm,file=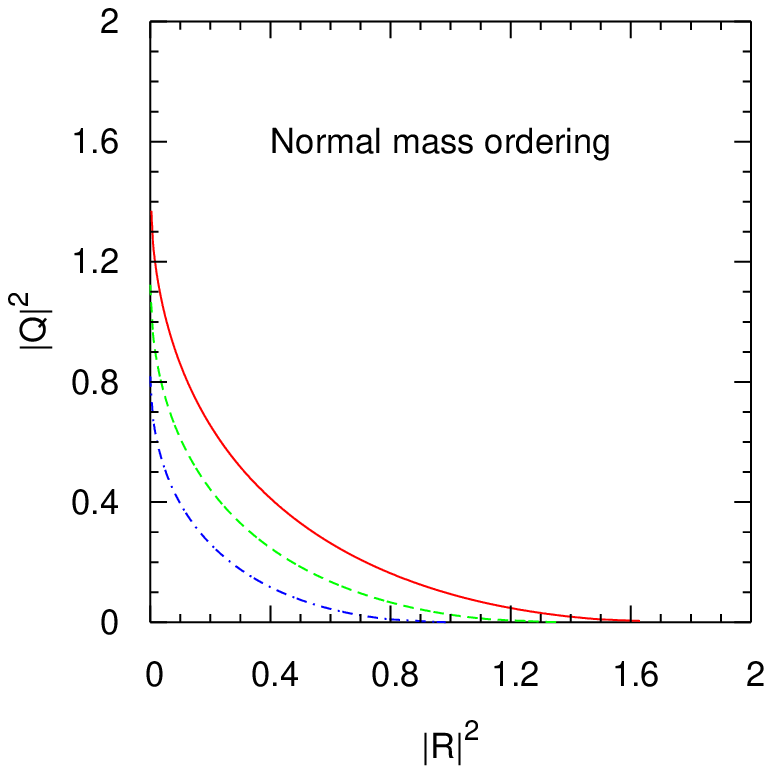}
\epsfig{width=7.0cm,file=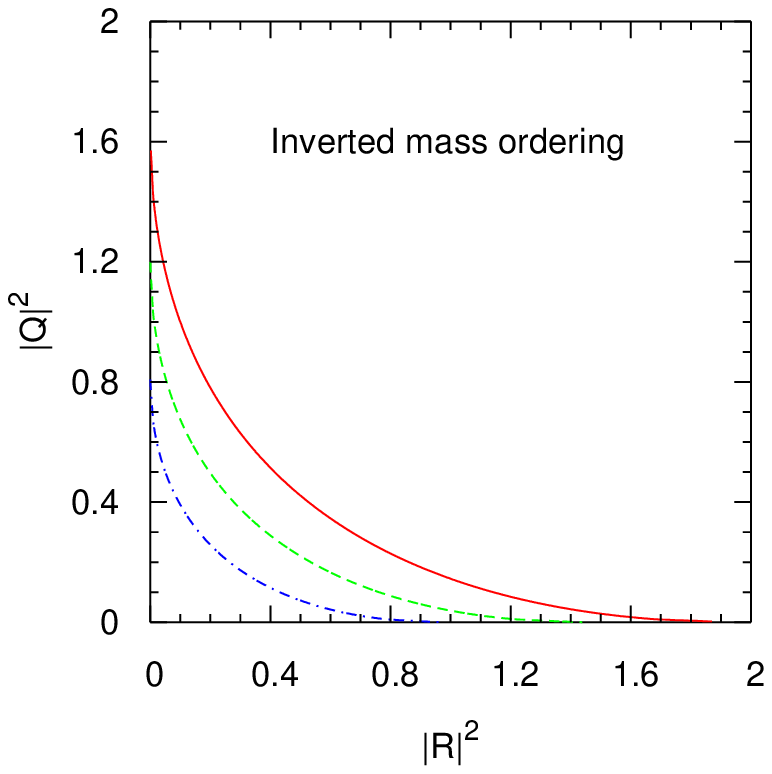}
\caption{Confidence level contours on $|Q|^2$--$|R|^2$ plane. 
The red (solid), 
green (dashed) and blue (dash-dotted) curves give the $3 \sigma$, 
$2 \sigma$ and $1 \sigma$ contours respectively. 
An energy independent error of $\pm 0.015$ on $P_{ee}$ 
has been taken into account.
We use the same $\dmsq_\odot$, 
$\theta_{12}$, $|\dma|$, $\theta_{23}$ and $\theta_{13}$ 
input values as in Fig.~\ref{Pmumu-fig1}. The additional input
values are $\delta_{cp}=\pi/3$ and $|Q|^2 = |R|^2 = 0$.
All the parameters other than $|Q|^2$ and $|R|^2$
are marginalized over in the analysis.}
\label{pee-contour}
\end{figure}

The magnitude of ${\cal A}_e (E)$ depends on 
$Q_1$, $Q_2$, $R_1$, $R_2$, $\delta_{cp}$ and mass ordering. 
To estimate the possibility of identifying any 
CPT violating signal in spite of our current lack of knowledge about 
the standard oscillation parameters in the CPT conserving case, 
we display the confidence level contours in 
Fig.~\ref{pee-contour}. 
We have chosen the best fit values of $\dmsq_\odot$, $\theta_{12}$, 
$\dma$, $\theta_{23}$ and $\theta_{13}$ \cite{lisi} as the input values.
Since we do not have any information 
about $\delta_{cp}$, we choose the input value of $\delta_{cp}$
in the range that is observed to give the most conservative
bound on $|Q|$ and $|R|$.
From the left panel 
of Fig.~\ref{pee-contour}, for the normal mass ordering the bounds obtained are
\beq 
|Q|^2 \lesssim 1.1 \; , \quad  
|R|^2 \lesssim 1.35 \;  \quad
(2\sigma) \; ,
\eeq
while the right panel with inverted mass ordering gives 
\beq 
|Q|^2 \lesssim 1.2 \; , \quad  
|R|^2 \lesssim 1.4 \;  \quad
(2\sigma) \; .
\eeq
It is observed that if the actual value of $\theta_{23}$ is smaller, 
the $|Q|^2$ bound decreases and the bound on $|R|^2$ becomes larger. 
The reverse is true when $\theta_{23}$ value is higher than the current 
best-fit value. 
This is true for both the mass orderings.

The bounds on $|Q|^2$ and $|R|^2$ translate to 
\begin{eqnarray}
|\mathbbm{H}_{b13}| & = & \epsilon  |Q| S_{E_0}
\lesssim  10^{-23} ~{\rm GeV} \; , \\
|\mathbbm{H}_{b12}| & = & \epsilon  |R| S_{E_0}\lesssim  
10^{-23} ~{\rm GeV} \; .
\label{bound-QR-CPTV}
\end{eqnarray}
The reach of ${\cal A}_e$ for the CPT violating observables 
is thus similar to that of ${\cal A}_\mu$ as obtained in
Sec.~\ref{sec-Pmumu}.
However, note that the actual combinations of elements of
${\bb H}_b$ constrained by the muon and electron channels
are quite different.


\section{Constraints from bounds on non-standard interactions}
\label{nsi}

In the presence of NSI of neutrinos with matter, the effective 
Hamiltonian in the three-flavor basis becomes
\beqa
{\bb H}_f \approx \mathbbm{U}_0 \cdot 
\frac{ {\rm diag}(0, \dmsq_{21}, \dmsq_{31})}{2E}  \cdot 
    \mathbbm{U}_0^\dagger
    + V_e \, \epsilon_{NSI}
    + {\rm diag}(V_e, 0, 0) \; ,
\label{Hf-2}
\eeqa
where $\epsilon_{NSI}$ is a $3\times 3$ matrix 
\beq
\epsilon_{NSI} = \left( \begin{array}{ccc}
\epsilon_{ee} & \epsilon_{e\mu} &  \epsilon_{e\tau} \\
\epsilon_{e \mu}^* & \epsilon_{\mu\mu} &  \epsilon_{\mu\tau} \\
\epsilon_{e \tau}^* & \epsilon_{\mu\tau}^* &  \epsilon_{\tau\tau} \\
\end{array} \right)
\label{def-epsilon}
\eeq
that parametrizes the NSI interactions. The factor of $V_e$
multiplying $\epsilon_{NSI}$ represents that the net 
NSI strength depends on the density of matter. 
The Hamiltonian for the antineutrinos will be obtained just by 
$V_e \to -V_e$ and 
$\epsilon_{\alpha \beta} \to \epsilon_{\alpha\beta}^\ast$.

Since CPT violation necessarily implies NSI, the bounds
on the NSI violating parameters $\epsilon_{\alpha \beta}$
would restrict CPT violating parameters as well. In order to
see the exact correspondence, note that the oscillation 
experiments are sensitive to only differences in the 
eigenvalues of the hamiltonian, and not to the absolute
eigenvalues. Therefore, the part of the NSI relevant for
oscillation experiments is only 
\beq
\mathbbm{H}_{NSI} \equiv \epsilon_{NSI}- \epsilon_{ee} \mathbbm{I} \; ,
\eeq
where $\mathbbm{I}$ is the identity matrix.
Then the comparison of eqs.~(\ref{Hf-1}) and (\ref{Hf-2})
implies that the mapping
\beq
\mathbbm{H}_b \Longleftrightarrow  \mathbbm{H}_{NSI} \; 
\label{map}
\eeq
would allow us to translate the results from one parametrization
to the other. Note that there is a difference between the two
sources of nonstandard physics under consideration.
Whereas $\mathbbm{H}_{NSI}$ is proportional to the matter density,
$\mathbbm{H}_b$ is independent of it.
However, as long as the matter density relevant for the experiments
restricting $\mathbbm{H}_{NSI}$ is known and is almost a constant,
the CPT violating contributions may be mimicked by the NSI ones.
Therefore, the bounds on $\epsilon_{\alpha\beta}$ from the NSI analysis 
can be translated to the bounds on the elements of
$\mathbbm{H}_b$  in the CPT parametrization.

Note that the bounds obtained from the
CPT analysis cannot be applied to the NSI bounds, since there
can be sources of NSI that are CPT conserving.
If an experiment 
is sensitive to the variations of
matter density along the neutrino path, it will be able to
separate the NSI contributions from the CPT violating ones.

A two flavor analysis of the atmospheric neutrino data
combined with the MACRO data \cite{maltoni-valle-atm}
and K2K data \cite{friedland-lunardini-atm} yields
\beq
-0.05 \lesssim \epsilon_{\mu\tau} \lesssim 0.04 \quad  (99\% ~{\rm C.L.})
\Rightarrow
|\mathbbm{H}_{b23}| \lesssim 10^{-23} ~{\rm GeV} \; ,
\eeq
where we have assumed an average density of $4.5$ g/cc inside the 
Earth. 
This bound is comparable to what would be obtained
using long baseline experiments as described in Sec.~\ref{sec-Pee}.

The neutrino scattering experiments CHARM and NuTeV 
mainly constrain the NSI couplings of $\nu_\mu$, and give 
\cite{sacha-davidson,charm,nutev}
\beq
|\epsilon_{e\mu}| \lesssim 10^{-3} \Rightarrow 
|\mathbbm{H}_{b23}| \lesssim 1.1 \times 10^{-25} ~ {\rm GeV} \; ,
\eeq
where we take the average Earth matter density to be
$2.7$ g/cc.
This constraint is extremely strong, and would imply
$|R| \approx 0$, thus simplifying the analysis of Sec.~\ref{sec-Pee}.
These experiments also bound
\beq
|\epsilon_{\mu\mu}| \lesssim 10^{-2} \; ,
\eeq
which by itelf does not put any constraints on the CPT violating
parameters since only the differences between the diagonal
elements of new physics hamiltonian is relevant.

Using the bounds on $\epsilon_{\mu \beta}$ stated above, 
\cite{yasuda-minos,friedland-lunardini-minos,ohlsson-minos} analyzed 
the possibility of constraing $\epsilon_{e e}$, $\epsilon_{e \tau}$ 
and $\epsilon_{\tau \tau}$ in MINOS experiment assuming 
$\epsilon_{e\mu} = \epsilon_{\mu\mu} = \epsilon_{\mu\tau} = 0$.
This effectively two-neutrino analysis leads to
\beq
|\epsilon_{e\tau}| \lesssim 2.9 \quad (99\% ~C.L.) \quad \Rightarrow \quad 
|\mathbbm{H}_{b13}| \lesssim 3.2 \times 10^{-22} ~{\rm GeV} \; ,
\eeq
which will be improved significantly at the neutrino factory
with the $\nu_e \to \nu_e$ channel, as described in Sec.~\ref{sec-Pee}.

The $99\%$ C. L. bounds on the diagonal NSI elements, 
given the initial assumption
of $\epsilon_{\mu\mu}=0$, translate as \cite{friedland-lunardini-minos}
\begin{eqnarray}
-0.4 \leq \epsilon_{\tau\tau} \leq 4.5  &  \Rightarrow &
-0.5 \times 10^{-22} ~{\rm GeV} 
< \mathbbm{H}_{b22} - \mathbbm{H}_{b33} < 5.0 \times 10^{-22} ~{\rm GeV} \\
-1.0 \leq \epsilon_{ee} \leq 0.9  &  \Rightarrow &
|\mathbbm{H}_{b22}| < 10^{-22} ~{\rm GeV} \; . 
\end{eqnarray}
Here, we have taken an average matter density of $2.7$ g/cc, which is
relevant for the MINOS baseline of 732 km. 
As seen in Sec.~\ref{sec-Pmumu}, NOvA will be able to constrain
$\mathbbm{H}_{b22} - \mathbbm{H}_{b33}$ to a much better accuracy.
The channels we have considered are rather insensitive to 
the absolute value of $|\mathbbm{H}_{b22}|$.

Current and future long baseline experiments like OPERA and T2KK
are expected to improve the bounds on NSI parameters
\cite{ohlsson-opera,nsi-t2kk}, 
and hence indirectly, those on the CPT violating parameters. 
Data from a future galactic supernova will also contribute to
constraints on NSI parameters \cite{ricard-nsi}, but 
converting them to bounds on CPT violation will not be
straightforward since the situation cannot be approximated
with a constant matter density.


\section{Summary and conclusions \label{conclusion}}


We have calculated possible CPT violating contributions to
neutrino masses and mixings in the complete three-flavor
analysis. 
Parametrizing the leading CPT violating
effects by a Hermitian matrix ${\bb b}$ that adds to the
effective neutrino Hamiltonian, we have developed a
framework based on the perturbative expansion in a small
auxiliary parameter $\epsilon \equiv 0.1$.
It involves expanding the elements of ${\bb H}_b$ (the matrix
${\bb b}$ in the flavor basis), the 
ratio $\dmsq_\odot/\dmsq_{atm}$, and $\theta_{13}$
as powers of $\epsilon$ multiplied by ${\cal O}(1)$
numbers. 
This allows us to treat the CPT violating ${\bb b}$
contributions in all generality, while keeping the
analytical expressions simple and transparent.
Though the complete parametrization of ${\bb b}$ involves 
three eigenvalues, three mixing angles and six phases,
we show that only certain combinations appear in the
survival probabilities of muon and electron neutrinos,
so that the analysis needs to concentrate only on
limiting those combinations.

The survival probabilities of $\nu_\mu$ and $\bar{\nu}_\mu$ 
to ${\cal O}(\epsilon)$
involve only two combinations of elements of ${\bb H}_b$,
{\it viz.} the real parameters $B_1 \propto 
\mathbbm{H}_{b22} - \mathbbm{H}_{b33}$ 
and $B_2 \propto  {\rm Re}(\mathbbm{H}_{b23})$.
Formally, the CPT violating contribution due to these terms 
is of a higher order than the CP violating contribution
in the CPT conserving limit.
The contribution due to $B_1$ vanishes when $\theta_{23}$
is maximal, so that a deviation of $\theta_{23}$ needs to be
established in order to put any bounds on this parameter.
The other quantity $B_2$ may be constrained to be 
$|B_2| \lesssim 0.1$ with 4 years of running with $\nu_\mu$
and $\bar{\nu}_\mu$ each at NOvA with an incident flux of 
$10^{21}$ pot yr$^{-1}$ at $2\sigma$.
This would correspond to bounds on 
$\mathbbm{H}_{b22} - \mathbbm{H}_{33}$ and ${\rm Re}(\mathbbm{H}_{b23})$
of the order $10^{-23}$ GeV.
Note that though the constraints that we have obtained are
of the same order as those obtained in earlier studies
\cite{barger-pakvasa,uma-sankar},
our analysis identifies the exact combination of elements
of ${\bb H}_b$, and hence ${\bb b}$, that these bounds apply to.

The CPT violating contribution to the survival probability of 
$\nu_e$ and $\bar{\nu}_e$ appears only at ${\cal O}(\epsilon^2)$,
and hence is expected to be more difficult to extract.
We isolate two different combinations of elements of ${\bb b}$,
{\it viz.} the complex parameters $Q \propto {\bb H}_{b13}$ 
and $R \propto {\bb H}_{b12}$, 
that govern this contribution and numerically analyze the
feasibility of extracting them. 
We demonstrate that for $|Q|, |R| \gtrsim 1.2$, it may be possible to
ascertain the presence of CPT violation at $2\sigma$ at a neutrino
factory with a detector at $L=3000$ km 
that can distinguish $\nu_e$ from $\bar{\nu}_e$,
within 4 years.
This corresponds to bounds on ${\bb H}_{b12}$ and 
${\bb H}_{b13}$ of the order $10^{-23}$ GeV.
Note that the exact combinations of elements of ${\bb H}_b$
that are constrained by the muon and electron channels
are quite different.

The CPT violating observables ${\cal A}_\mu$ and ${\cal A}_e$
in this paper are the same as those considered in 
\cite{amol-shamayita} for disentangling the signals of
sterile neutrinos. The energy dependence of the
signatures of CPT violation and sterile neutrinos, however,
is different and these two new physics signatures may
be disentangled with a combined analysis.

The constraints obtained on the NSI parameters through
oscillation and non-oscillation experiments can be translated
to bounds on elements of ${\bb H}_b$. We find that the 
bound on $|{\bb H}_{b12}|$ implied by the NSI constraints
is much stronger than the expected reach of even neutrino factories, 
whereas the bound on $|{\bb H}_{b23}|$ is comparable to
the one expected at NOvA.
On the other hand, $|{\bb H}_{b13}|$ and the difference
${\bb H}_{b22}-{\bb H}_{b33}$ will be much better constrained by the
long baseline experiments.
NSI analyses give a constraint on the absolute value of
${\bb H}_{b22}$, to which
the channels we have considered are rather insensitive.

In this paper, we have confined ourselves to low energies
($E < 5$ GeV) for the muon channel and high energies
($E> 15$ GeV) for the electron channel.
This allowed us to cleanly isolate certain combinations of 
elements of ${\bb H}_b$, {\it viz.} two real quantities 
${\bb H}_{b22}-{\bb H}_{b33}, {\rm Re}({\bb H}_{b23})$
through the muons and two complex quantities 
${\bb H}_{b12}, {\bb H}_{b13}$ through the electrons.
A more exhaustive analysis that uses the complete energy range 
and the long baseline as well as the atmospheric neutrino data 
may lead to constraints on other combinations of elements of 
${\bb H}_b$. 
However, it is not clear if it can be achieved through a clean
analytic treatment.


\section*{Acknowledgements}

We would like to thank P. Huber, T. Schwetz and W. Winter for
their guidance in using GLoBES
and S. Uma Sankar for helpful comments on the manuscript. 
This work was partly supported through the
Partner Group program between the Max Planck Institute
for Physics and Tata Institute of Fundamental Research.



\begin{thebibliography}{99}

\bibitem{coleman-glashow}
  S.~R.~Coleman and S.~L.~Glashow,
  Phys.\ Rev.\  D {\bf 59}, 116008 (1999)
  [arXiv:hep-ph/9812418].

\bibitem{greenberg-1} 
  O.~W.~Greenberg,
  Phys.\ Rev.\ Lett.\  {\bf 89}, 231602 (2002)
  [arXiv:hep-ph/0201258].

\bibitem{greenberg-2}
  O.~W.~Greenberg,
  Found.\ Phys.\  {\bf 36}, 1535 (2006)
  [arXiv:hep-ph/0309309].



\bibitem{kostelecky}
  D.~Colladay and V.~A.~Kostelecky,
  Phys.\ Rev.\  D {\bf 55}, 6760 (1997)
  [arXiv:hep-ph/9703464].




\bibitem{mattingly}
  D.~Mattingly,
  Living Rev.\ Rel.\  {\bf 8}, 5 (2005)
  [arXiv:gr-qc/0502097].

\bibitem{arkani-hamed}
  N.~Arkani-Hamed, S.~Dimopoulos, G.~R.~Dvali and J.~March-Russell,
  Phys.\ Rev.\  D {\bf 65}, 024032 (2001)
  [arXiv:hep-ph/9811448].

\bibitem{huber}
  S.~J.~Huber and Q.~Shafi,
  Phys.\ Lett.\  B {\bf 512}, 365 (2001)
  [arXiv:hep-ph/0104293].

\bibitem{grossman}
  Y.~Grossman and M.~Neubert,
  Phys.\ Lett.\  B {\bf 474}, 361 (2000)
  [arXiv:hep-ph/9912408].

\bibitem{barenboim}
  G.~Barenboim and J.~D.~Lykken,
  Phys.\ Lett.\  B {\bf 554}, 73 (2003)
  [arXiv:hep-ph/0210411].

\bibitem{k-kbar} 
 W.~M.~Yao {\it et al.}  [Particle Data Group],
  J.\ Phys.\ G {\bf 33}, 1 (2006).


\bibitem{Bd-Bdbar}
  C.~Leonidopoulos  [BELLE collaboration],
  arXiv:hep-ex/0107001.



\bibitem{anomalous-g} 
 G.~W.~Bennett {\it et al.}  [Muon (g-2) Collaboration],
   Phys.\ Rev.\ Lett.\  {\bf 100}, 091602 (2008)
   [arXiv:0709.4670 [hep-ex]].



\bibitem{LSND}
  A.~Aguilar {\it et al.}  [LSND Collaboration],
  Phys.\ Rev.\  D {\bf 64}, 112007 (2001)
  [arXiv:hep-ex/0104049].


\bibitem{LSND-Smirnov} 
  G.~Barenboim, L.~Borissov, J.~D.~Lykken and A.~Y.~Smirnov,
  JHEP {\bf 0210}, 001 (2002)
  [arXiv:hep-ph/0108199].

\bibitem{Yanagida}
  H.~Murayama and T.~Yanagida,
  Phys.\ Lett.\  B {\bf 520}, 263 (2001)
  [arXiv:hep-ph/0010178].

\bibitem{gouvea-cpt}
  A.~De Gouvea,
  Phys.\ Rev.\  D {\bf 66}, 076005 (2002)
  [arXiv:hep-ph/0204077].



\bibitem{kamland}
  J.~Shirai  [KamLAND Collaboration],
  Nucl.\ Phys.\ Proc.\ Suppl.\  {\bf 168}, 77 (2007).

\bibitem{MiniBooNE}
  A.~A.~Aguilar-Arevalo {\it et al.}  [The MiniBooNE Collaboration],
  Phys.\ Rev.\ Lett.\  {\bf 98}, 231801 (2007)
  [arXiv:0704.1500 [hep-ex]].


\bibitem{lisi}
  G.~L.~Fogli {\it et al.},
  Nucl.\ Phys.\ Proc.\ Suppl.\  {\bf 168}, 341 (2007).


\bibitem{barger-pakvasa} 
  V.~D.~Barger, S.~Pakvasa, T.~J.~Weiler and K.~Whisnant,
  Phys.\ Rev.\ Lett.\  {\bf 85}, 5055 (2000)
  [arXiv:hep-ph/0005197].


\bibitem{uma-sankar}
  A.~Datta, R.~Gandhi, P.~Mehta and S.~Uma Sankar,
  Phys.\ Lett.\  B {\bf 597}, 356 (2004)
  [arXiv:hep-ph/0312027].

\bibitem{solar-KamLAND}
  J.~N.~Bahcall, V.~Barger and D.~Marfatia,
  Phys.\ Lett.\  B {\bf 534}, 120 (2002)
  [arXiv:hep-ph/0201211].


\bibitem{bilenky}
  S.~M.~Bilenky, M.~Freund, M.~Lindner, T.~Ohlsson and W.~Winter,
  Phys.\ Rev.\  D {\bf 65}, 073024 (2002)
  [arXiv:hep-ph/0112226].

\bibitem{garcia-maltoni}
  M.~C.~Gonzalez-Garcia and M.~Maltoni,
  Phys.\ Rev.\  D {\bf 70}, 033010 (2004)
  [arXiv:hep-ph/0404085].


\bibitem{maltoni-valle-atm}
  N.~Fornengo, M.~Maltoni, R.~T.~Bayo and J.~W.~F.~Valle,
  Phys.\ Rev.\  D {\bf 65}, 013010 (2001)
  [arXiv:hep-ph/0108043].

\bibitem{friedland-lunardini-atm}
  A.~Friedland and C.~Lunardini,
  Phys.\ Rev.\  D {\bf 72}, 053009 (2005)
  [arXiv:hep-ph/0506143].

\bibitem{sacha-davidson}
  S.~Davidson, C.~Pena-Garay, N.~Rius and A.~Santamaria,
  JHEP {\bf 0303}, 011 (2003)
  [arXiv:hep-ph/0302093].

\bibitem{yasuda-minos}
  N.~Kitazawa, H.~Sugiyama and O.~Yasuda,
  arXiv:hep-ph/0606013.

\bibitem{friedland-lunardini-minos}
  A.~Friedland and C.~Lunardini,
  Phys.\ Rev.\  D {\bf 74}, 033012 (2006)
  [arXiv:hep-ph/0606101].

\bibitem{ohlsson-minos}
  M.~Blennow, T.~Ohlsson and J.~Skrotzki,
  Phys.\ Lett.\  B {\bf 660}, 522 (2008)
  [arXiv:hep-ph/0702059].



\bibitem{nova-1} 
  D.~S.~Ayres {\it et al.}  [NOvA Collaboration],
  arXiv:hep-ex/0503053.

\bibitem{nova-2}
T. ~Yang and S. ~Wojcicki  [NOvA Collaboration],
 Off-Axis-Note-SIM-30 (2004).

\bibitem{prem}
  A.~M.~Dziewonski and D.~L.~Anderson,
  Phys.\ Earth Planet.\ Interiors {\bf 25}, 297 (1981).

\bibitem{globes-1}
  P.~Huber, M.~Lindner and W.~Winter,
  Comput.\ Phys.\ Commun.\  {\bf 167}, 195 (2005)
  [arXiv:hep-ph/0407333].

\bibitem{globes-2}
  P.~Huber, J.~Kopp, M.~Lindner, M.~Rolinec and W.~Winter,
  Comput.\ Phys.\ Commun.\  {\bf 177}, 432 (2007)
  [arXiv:hep-ph/0701187].

\bibitem{xsec-1}
  M.~D.~Messier, UMI-99-23965.

\bibitem{xsec-2}
  E.~A.~Paschos and J.~Y.~Yu,
  Phys.\ Rev.\  D {\bf 65}, 033002 (2002)
  [arXiv:hep-ph/0107261].

\bibitem{maximal-t23} 
  S.~Choubey and P.~Roy,
  Phys.\ Rev.\  D {\bf 73}, 013006 (2006)
  [arXiv:hep-ph/0509197].

\bibitem{nufact-setup} 
  C.~H.~Albright {\it et al.}  [Neutrino Factory/Muon Collider
                  Collaboration],
  arXiv:physics/0411123.

\bibitem{nufactII}
  P.~Huber, M.~Lindner and W.~Winter,
  Nucl.\ Phys.\  B {\bf 645}, 3 (2002)
  [arXiv:hep-ph/0204352].


\bibitem{charm}
  P.~Vilain {\it et al.}  [CHARM-II Collaboration],
  Phys.\ Lett.\  B {\bf 335}, 246 (1994).

\bibitem{nutev}
  G.~P.~Zeller {\it et al.}  [NuTeV Collaboration],
  Phys.\ Rev.\ Lett.\  {\bf 88}, 091802 (2002)
  [arXiv:hep-ex/0110059].


\bibitem{ohlsson-opera}
 M.~Blennow, D.~Meloni, T.~Ohlsson, F.~Terranova and M.~Westerberg,
  arXiv:0804.2744 [hep-ph].

\bibitem{nsi-t2kk}
  H.~Minakata,
  arXiv:0805.2435 [hep-ph].

\bibitem{ricard-nsi}
A.~Esteban-Pretel, R.~Tomas and J.~W.~F.~Valle,
  Phys.\ Rev.\  D {\bf 76}, 053001 (2007)
  [arXiv:0704.0032 [hep-ph]].


\bibitem{amol-shamayita}
  A.~Dighe and S.~Ray,
  Phys.\ Rev.\  D {\bf 76}, 113001 (2007)
  [arXiv:0709.0383 [hep-ph]].


\end{thebibliography}
\end{document}